\documentclass[12pt, onecolumn]{IEEEtran}   
\usepackage{setspace}
\usepackage{mathptmx}                    
\usepackage[margin=1in]{geometry}        
\usepackage{amsmath}
\usepackage{cite}
\usepackage{amsfonts}
\usepackage{graphicx}
\usepackage{xcolor}
\usepackage{booktabs}
\usepackage{subcaption}
\usepackage{listings}
\usepackage{caption}
\usepackage{algorithm}   
\usepackage{algpseudocode} 

\ifCLASSINFOpdf
\else
\fi

\begin{document}
%
\title{Feature Partitioning and Semantic Equalization for Intrinsic Robustness in Semantic Communication under Packet Loss}
%
%
%

\author{Xiao~Yang,
        Shuai~Ma,
        Yong~Liang,
        and~Guangming~Shi,~\IEEEmembership{Fellow,~IEEE}

\thanks{Xiao Yang is with the Shenzhen Institutes of Advanced Technology,
Chinese Academy of Sciences, Shenzhen 518055, China, also with Pengcheng Laboratory, Shenzhen 518066, China, and also with University of Chinese
Academy of Sciences, Beijing 100049, China. (e-mail: yangx20@pcl.ac.cn).}
\thanks{Shuai Ma, Yong Liang are with Pengcheng Laboratory, Shenzhen 518066, China. (e-mail:
\{mash01, liangy02\}@pcl.ac.cn).}
\thanks{Guangming Shi is with the School of Artificial Intelligence, Xidian University, Xi’an, Shaanxi 710071, China, and also with Pengcheng Laboratory,
Shenzhen, 518066, China (e-mail: gmshi@xidian.edu.cn).}
}

\maketitle

\begin{abstract}

Semantic communication can improve transmission efficiency by focusing on task-relevant information. However, under packet-based communication protocols, any error typically results in the loss of an entire packet, making semantic communication particularly vulnerable to packet loss. Since high-dimensional semantic features must be partitioned into 1-D transmission units during packetization. A critical open question is how to partition semantic features to maximize robustness. To address this, we systematically investigate the performance of two mainstream architectures, Transformer and Convolutional neural networks (CNNs), under various feature partitioning schemes. The results show that the Transformer architecture exhibits inherent robustness to packet loss when partitioned along the channel dimension. In contrast, the CNNs-based baseline exhibits imbalanced channel utilization, causing severe degradation once dominant channels are lost. To enhance the CNNs’ resilience, we propose a lightweight Semantic Equalization Mechanism (SEM) that balances channel contributions and prevents a few channels from dominating. SEM consists of two parallel approaches: a Dynamic Scale module that adaptively adjusts channel importance, and a Broadcast module that facilitates information interaction among channels. Experimental results demonstrate that CNNs equipped with SEM achieve graceful degradation under packet loss (retaining about 85\% of lossless PSNR at 40\% packet loss), comparable to that of Transformer models. Our findings indicate that, under an appropriate partitioning strategy, maintaining a balanced semantic representation is a fundamental condition for achieving intrinsic robustness against packet loss. These insights may also extend to other modalities such as video and support practical semantic communication design.
\end{abstract}


%
\IEEEpeerreviewmaketitle

\section{Introduction}
%
%
%
%
\IEEEPARstart{A}{s} an emerging communication paradigm, semantic communication~(SemCom) demonstrates significant potential in transmission efficiency and compression performance compared to traditional communication systems by extracting and transmitting semantic features~\cite{chaccour2024less, getu2025semantic}. While traditional communication relies on the lossless transmission of bit-level information, SemCom focuses on the representation of task-related or reconstruction-related information, significantly improving compression ratios and reducing transmission bandwidth. SemCom is expected to become a groundbreaking communication technology in next-generation communication systems, particularly for applications involving multimedia, Internet-of-Things (IoT), and real-time collaborative intelligence~\cite{zhang2024intellicise}.

Building upon these advantages, a growing body of research has explored the design and optimization of SemCom systems. \cite{xie2021deepsc} establishes a text-modal SemCom system named DeepSC based on the transformer architecture, aiming to recover the meaning of sentences rather than restoring bit-level information. This system demonstrates robustness to channel variations and achieves superior performance under low signal-to-noise ratio (SNR) conditions. SwinJSCC \cite{yang2024swinjscc} adopts the swin transformer \cite{swin} instead of convolutional neural n
networks (CNNs)~\cite{ipc2019TCCN} as the backbone for image SemCom and incorporates two modulation modules based on channel state information and transmission rates. Experimental results show that SwinJSCC achieves comparable performance to the existing BPG+LDPC communication scheme. For the video modality, the authors of \cite{wang2022dvst} propose a SemCom framework designed to minimize end-to-end rate-distortion performance. Experiments demonstrate that it outperforms traditional video coding schemes and achieves excellent performance in visual tasks. However, most of these rely on analog transmission.

To ensure compatibility with established digital communication systems, the paradigm has gradually evolved toward digital SemCom, where semantic features are encoded into discrete representations compatible with standard transmission protocols. Key enabling techniques include quantization \cite{huang2025d, lyu2025vq, shen2025semantic, liu2024ofdm, gong2025digital}—either by codebook-based vector quantization or integer-level discretization—and joint coding-modulation \cite{bo2024joint, zhang2024analog}, among others.

\begin{figure*}[htbp]
    \centering
    \includegraphics[width=0.95\textwidth]{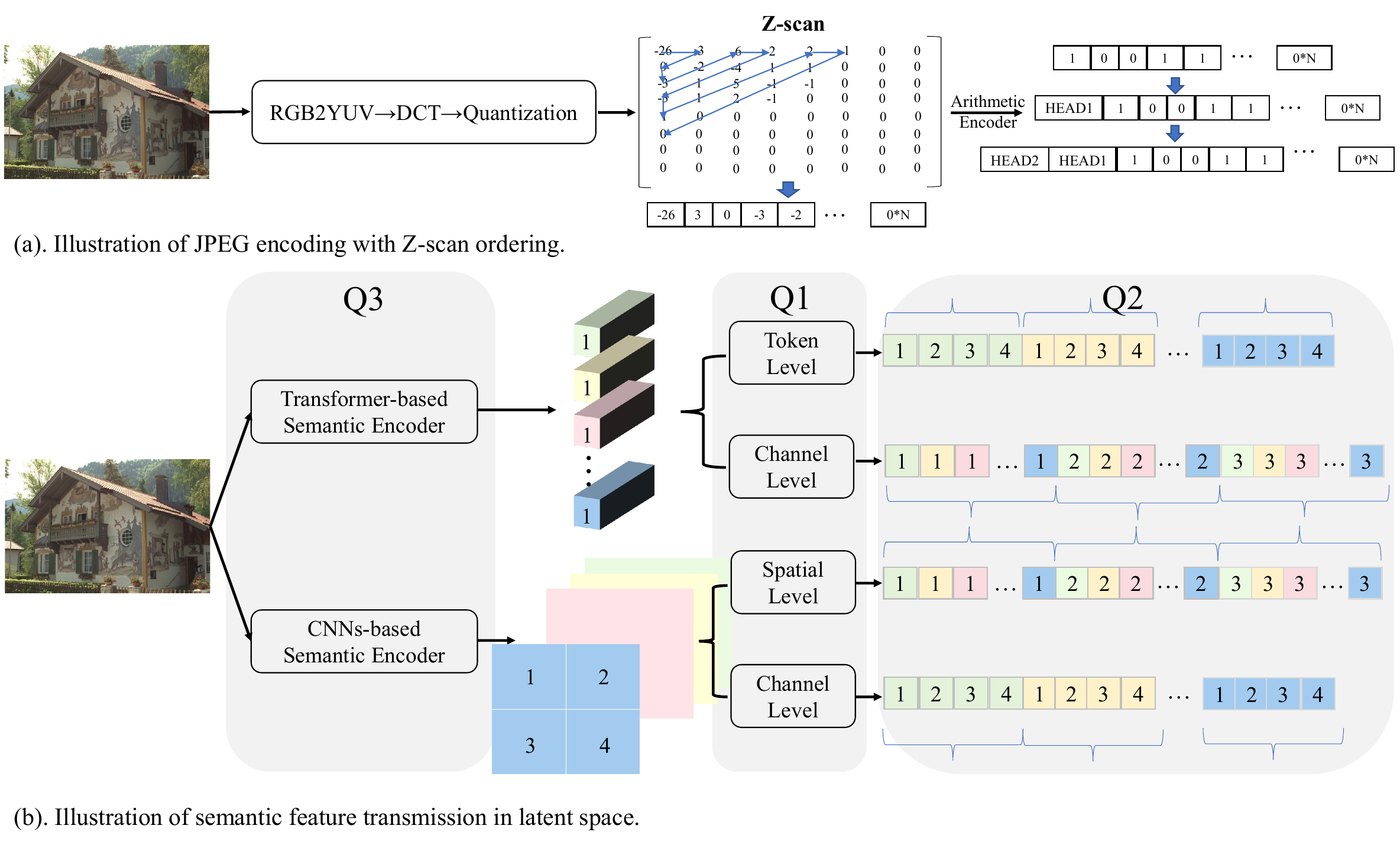}
    \caption{Partition strategies from JPEG to semantic communication.}
    \label{fig:part1_shift}
\end{figure*}

Beyond semantic transmission at the physical layer, packet-based communication at the network layer introduces additional challenges. In particular, high-dimensional semantic features need to be segmented and encapsulated into packets for transmission. As a result, the loss of a single packet may lead to the complete loss of all semantic information contained within it, severely degrading reconstruction quality and transmission reliability. Addressing this issue is therefore critical and constitutes the focus of this work. 

Recent studies \cite{wang2025resicomp, teng2025conquering} have begun to explore this problem, proposing various strategies to enhance resilience against packet loss. Specifically, Wang et al.~\cite{wang2025resicomp} propose ResiComp which employs masked modeling during training to simulate packet loss, and offers flexible coding modes to balance efficiency and resilience. \cite{teng2025conquering} proposes MSTVSC, a packet-loss-resistant video SemCom system. The approach employs semantic-level interleaving and a 3D CNNs at the receiver to recover missing information caused by packet drops. Simulations show that MSTVSC achieves strong reconstruction performance even under high packet loss conditions.

However, these approaches primarily focus on predicting lost data rather than considering how semantic information should be organized and transmitted. Unlike conventional bit streams, semantic features encode distributed and correlated information, typically represented as high-dimensional latent vectors in which each dimension carries distinct semantic content.

As illustrated in Fig.~\ref{fig:part1_shift}, traditional codecs such as JPEG convert a two-dimensional coefficient map into a one-dimensional sequence via zig-zag scanning, whereas semantic communication maintains the multi-dimensional structure of latent features during transmission, preserving the inherent semantic correlations.
This raises a series of fundamental questions: 
\begin{itemize}
    \item Q1: How should the encoded semantic features be partitioned into transmission units?
    \item Q2: Does the choice of partitioning dimension—such as channel-level, token-level, or intra-token segmentation—affect the system’s robustness against packet loss? 
    \item Q3: Is this relationship dependent on the backbone network architecture?
\end{itemize}

To address these questions, this work investigates the structural organization of semantic features from the perspective of their partitioning and transmission.
Through extensive experiments, we systematically evaluate how different partitioning strategies—such as channel-wise, token-wise, and intra-token segmentation—affect the robustness of SemCom systems under packet loss across two mainstream architectures, CNNs and Transformers.
Building upon these empirical findings, we design a lightweight semantic equalization mechanism (SEM) that balances the importance of semantic features, enabling graceful degradation and enhanced resilience during packet loss.

The main contributions of this paper are summarized as follows:
\begin{itemize}
    \item \textbf{We systematically examine how different feature partitioning strategies affect the robustness of semantic communication systems.} Extensive experiments are conducted on both CNNs- and Transformer-based systems, evaluating multiple partitioning approaches, including channel-wise, token-wise, and spatial-token segmentation, to analyze their impact on system resilience under packet loss.

    The results indicate that different partitioning strategies lead to varying levels of robustness under packet loss (Q2) and that CNNs- and Transformer-based architectures exhibit distinct resilience characteristics (Q3). Overall, partitioning features along the channel dimension proves to be a consistently effective strategy for both architectures (Q1). Specifically, Transformers maintain stable reconstruction quality even when some channels are lost, whereas CNNs display an uneven distribution of channel importance: they achieve good reconstruction quality when non-critical channels are lost, but suffer significant degradation when key channels are affected.
    \item \textbf{We propose a lightweight semantic equalization mechanism to enable graceful degradation under packet loss channel.} Building upon the observation that CNNs-based SemCom systems exhibit an uneven distribution of channel importance, SEM is proposed to balance the contribution of each channel. The mechanism consists of two parallel approaches:
(1) a dynamic scaling strategy, which adjusts the weights of different channels to redistribute feature importance; and
(2) a broadcast strategy, which treats each channel as a node and diffuses information across channels through feature broadcasting.

The entire mechanism introduces either no additional parameters or less than 1\% of the original model parameters, and it can be seamlessly trained in an end-to-end manner without extra fine-tuning. Beyong this, the proposed approach may also be extended to task-oriented communication and other scenarios where feature or task distributions are inherently imbalanced.
    \item \textbf{Comprehensive experiments are conducted to validate the effectiveness of the proposed mechanism.} Experimental results on image transmission tasks demonstrate that the proposed SEM effectively enables CNNs-based SemCom systems to achieve smoother performance degradation under packet loss, rather than abrupt quality collapse. In addition, we also provide a new perspective for understanding the robustness of SemCom systems against packet loss, highlighting how balanced feature representations inherently enhance resistance to degradation.
\end{itemize}

The rest of this paper is organized as follows. The system model and objectives is introduced in Section \ref{sys_model}. The analysis of partitioning strategies on semantic representations is presented in Section III. The semantic equalization mechanism is introduced in Section IV. Finally, experimental results are presented in Section V, and conclusions are drawn in Section VI.

\section{SYSTEM MODEL AND OBJECTIVES}
\label{sys_model}

\subsection{Semantic Communication System Model Based on UDP}

\begin{figure*}[htbp]
    \centering
    \includegraphics[width=0.99\textwidth]{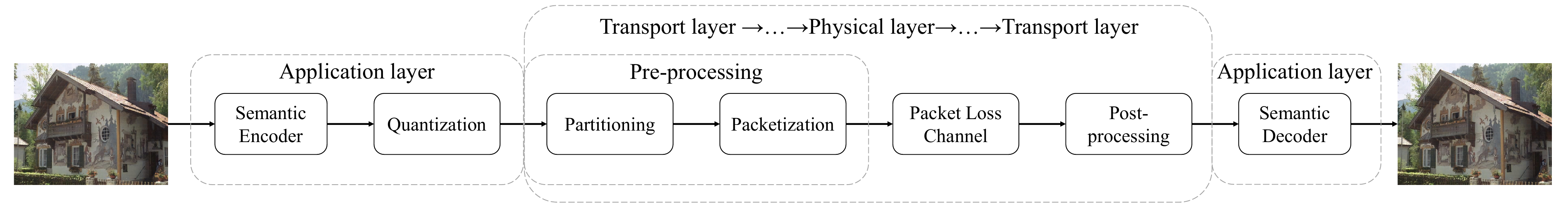}
    \caption{Semantic communicaiton system model based on UDP.}
    \label{fig:sys_demo}
\end{figure*}

For simplicity, we consider a typical SemCom system consisting of four main components: a semantic encoder, a quantization module, a packet-loss channel, and a semantic decoder, as illustrated in Fig.~\ref{fig:sys_demo}. 

In the design of the transport layer, we adopt the  UDP (User Datagram Protocol) protocol due to its low latency and lightweight nature. Unlike TCP (Transmission Control Protocol), UDP does not provide retransmission or congestion control mechanisms; once a packet is lost, it cannot be recovered through feedback \footnote{Here, we do not elaborate on the design of other communication layers, since packet loss primarily affects the transport layer in terms of packet organization and transmission strategy, rather than the signal characteristics of the physical or link layers. Our focus is on enhancing overall semantic robustness under given channel conditions.}.

The function of each module is described as follows:
\begin{itemize}
    \item \textbf{Semantic Encoder}: The encoder maps and downsamples input data (e.g., an image $x \in \mathbb{R}^{H \times W \times C}$) into high-dimensional vectors, which depends on the network architecture. When the encoder is based on CNNs, the original dimensions are mapped to $y \in \mathbb{R}^{c \times h \times w}$, while a transformer architecture maps to $y \in \mathbb{R}^{n \times c}$. Additionally, these features are typically continuous-valued vectors.

    \item \textbf{Quantization}: To enable digital transmission, semantic features must be quantized into discrete representations $\bar{y}$. We adopt a uniform scalar quantizer~\cite{balle2016end}. During training, quantization is approximated by injecting uniform noise $\mathcal{U}(-0.5,0.5)$:
    \[
    \bar{y} = y + \epsilon, \quad \epsilon \sim \mathcal{U}(-0.5, 0.5),
    \]
    which provides a differentiable approximation for backpropagation. During testing, quantization is implemented by rounding to the nearest integer:
    \[
    \bar{y} = \text{round}(y).
    \]
    The quantified features are then divided into a set of semantic units.
    \[
    \left\{z_i\right\}_{i=1}^N=\mathcal{P}(\bar{y}), 
    \]
    where each $z_i$ represents a transmission unit, i.e., the minimal packetized feature block sent through the channel.

    \item \textbf{Packet-loss Channel}: During transmission, packets carrying semantic units may be randomly dropped due to the unreliable nature of UDP. We model this as a \emph{random erasure channel}, where each packet is erased with probability $p$~\footnote{$p$ is affected by various factors such as SNR, packet length, and network conditions. 
Modeling the detailed relationship between these factors and the loss probability is beyond the scope of this work, and we directly focus on the resulting packet loss rate.}
. Following traditional communication design, we assume that the semantic features of a packet are either entirely received or completely lost.
    \[ \hat{z}_i= \begin{cases} z_i, & \text{w.p. } 1-p, \\ 0, & \text{w.p. } p . \end{cases} \]

    \item \textbf{Semantic Decoder}: The receiving end first aggregates the possibly incomplete set of received semantic units $\hat{z}_i$ into a feature representation $\hat{y}$ and then decodes it to produce the final output $\hat{x}$. 
\end{itemize}

Formally, the end-to-end SemCom process can be expressed as the following markov chain:

\[
x 
\;\xrightarrow{Enc}\;
y 
\;\xrightarrow{}\;
\bar{y}
\;\xrightarrow{\mathcal{P}}\;
\{z_i\}_{i=1}^{N}
\;\xrightarrow{Ch}\;
\{\hat{z}_i\}_{i=1}^{N}
\;\xrightarrow{\mathcal{A}}\;
\hat{y}
\;\xrightarrow{Dec}\;
\hat{x}.
\]

This UDP-based abstraction captures the essence of SemCom: transmitting meaningful features through lightweight transport while tolerating packet loss.

\subsection{Preliminaries}
This part first explains the motivation for choosing the CNNs and Transformer architectures, and then introduces each architecture along with its corresponding feature partitioning strategies.

\paragraph{Architecture Selection}
With the rapid development of deep learning, a variety of network architectures have been proposed—such as transformers~\cite{transformer}, CNNs, KANs~\cite{liu2024kan}, and Mamba~\cite{gu2024mamba} networks—many of which have been applied to SemCom systems~\cite{wu2024mambajscc, yang2024swinjscc, ipc2019TCCN, wang2024swin}. Among them, transformers and CNNs stand out as the most representative and widely adopted architectures. Transformers are distinguished by their ability to model global dependencies through self-attention mechanisms and their strong parallel computing capability, which enables efficient processing of long sequences and large-scale data. In contrast, CNNs excel at capturing local spatial correlations via convolutional operations and are highly optimized for hardware implementation, benefiting from mature acceleration frameworks and efficient memory usage.

To clearly demonstrate and validate our proposed framework, we select CNNs and transformers as representative backbones for analysis and experimentation. Nevertheless, the methodology presented in this work is not limited to these two architectures and can be readily extended to other network structures in future research.
\paragraph{Transformer tokens.}  
In transformer-based models such as ViT and Swin transformer, the input image $x \in \mathbb{R}^{C \times H \times W}$ is partitioned into non-overlapping patches, which are then projected into a sequence of $N$ semantic tokens:
\[
T = [t_1, t_2, \ldots, t_N], \quad t_i \in \mathbb{R}^d ,
\]
where $d$ denotes the token dimension. Each token $t_i$ initially encodes local spatial semantics of the corresponding image patch. Through the multi-head self-attention mechanism, however, tokens iteratively exchange information, enabling each $t_i$ to integrate not only its local content but also long-range contextual dependencies across the entire image. As a result, tokens simultaneously preserve fine-grained local structure and capture global semantic correlations, making them the fundamental representation and transmission units in transformer-based SemCom systems.

From the partitioning perspective, the encoded semantic features can be organized and transmitted in different granularities.
They can be sent as complete tokens, referred to as \emph{inter-token partitioning}, or further divided into sub-vectors along their $d$-dimensional feature space, referred to as \emph{inter-channel partitioning}.

In inter-token partitioning, each token serves as an independent semantic unit, and packet loss corresponds to the complete loss of certain tokens. In channel-wise partitioning, each semantic unit corresponds to a specific feature dimension across all tokens, which means packet loss removes that particular dimension for every token.

\paragraph{CNNs channels and spatial features.}  
In CNNs-based SemCom systems, the bottleneck feature representation after the encoder is expressed as a tensor
\[
y \in \mathbb{R}^{c \times h \times w},
\]
where $c$ denotes the number of channels, and $h$ and $w$ represent the spatial resolution. Each channel encodes a distinct semantic abstraction of the input, while spatial dimensions preserve the structural arrangement of features.

Building on the idea that convolutional features can be adaptively emphasized along different dimensions, as demonstrated in SENet \cite{hu2018squeeze} for channel attention and CBAM~\cite{woo2018cbam} for joint channel--spatial attention, we design two semantic unit partitioning strategies: channel-based partitioning and spatial-based partitioning.

For transmission, these two strategies can be interpreted from distinct feature partitioning perspectives:

\begin{itemize}
    \item \emph{Channel-wise partitioning:} Each unit corresponds to a specific set of feature channels across the entire spatial map.
    Packet loss therefore removes the semantic information associated with those channels throughout the representation.
    \item \emph{Spatial-based partitioning:} Each unit corresponds to a spatial region within the feature map, containing all channels at that location.  
Packet loss thus causes localized spatial erasures while preserving information from other regions.
\end{itemize}

\subsection{Random Grouping}

Considering packet length constraints and transmission efficiency, multiple semantic units ${z_i}$ are typically combined into a single packet before transmission.
A straightforward approach is to arrange these units in sequential order; however, this design is highly sensitive to burst losses, as consecutive packets may carry semantically adjacent units, resulting in correlated information loss and degraded reconstruction quality.

To mitigate this issue, we introduce a random unit grouping mechanism at the encoder, which randomly mixes semantic units prior to transmission.
This strategy functions analogously to random interleaving in~\cite{teng2025conquering}. The detailed procedure of the random unit grouping mechanism is presented as follows.

\begin{figure}[htbp]
    \centering
    \includegraphics[width=0.88\linewidth]{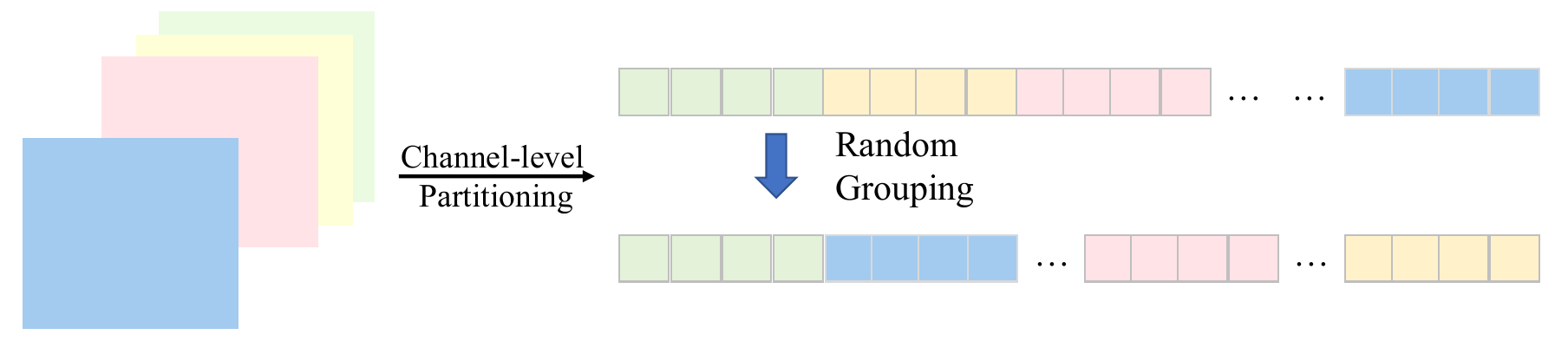}
    \caption{Random grouping: illustration of the interleaving mechanism that disperses consecutive semantic units into different transmission packets, transforming burst losses into approximately independent errors.}
    \label{fig:random_group}
\end{figure}

\begin{enumerate}
\item After generating semantic units $\{z_i\}_{i=1}^{N}$, we randomly group them into multiple transmission units according to a permutation $\pi$.
\item At the receiver, the inverse mapping $\pi^{-1}$ is applied to restore the original grouping before decoding.
\end{enumerate}

This operation preserves the original information content while redistributing consecutive losses across distinct feature groups, effectively transforming burst losses into approximately independent ones and enhancing robustness against correlated packet loss.

\subsection{Robustness Objective}
We consider image transmission tasks, where the goal is to minimize reconstruction distortion or maximize perceptual quality. Typical evaluation metrics include Mean Squared Error (MSE), Peak Signal-to-Noise Ratio (PSNR), and Structural Similarity Index Measure (SSIM).

Since the packet loss rate $p$ is typically random and uncontrollable, a robust semantic communication system is not expected to maintain perfectly stable PSNR/SSIM under all conditions. Instead, it should achieve \emph{graceful degradation}, where reconstruction quality decreases smoothly as $p$ increases rather than collapsing abruptly. Importantly, such robustness should be attained without relying on additional priors or auxiliary networks.

In summary, a robust SemCom system should satisfy:
\begin{itemize}
\item \textbf{Graceful degradation}: reconstruction quality decreases smoothly with increasing $p$.
\item \textbf{High-loss tolerance}: performance remains acceptable even under severe packet loss.
\end{itemize}


\section{Analysis of Partitioning Strategies on Semantic Representations}
In this section, we investigate how different semantic features partitioning strategies affect the robustness of semantic communication systems, focusing on both Transformer- and CNNs-based architectures in image transmission tasks.
Both architectures encode visual information into intermediate feature maps or token sequences, which can be divided along various dimensions, such as spatial, channel, or token axes.
\subsection{Feature Partitioning in Transformer-based Semantic Communication Systems}

\paragraph{Model Selection and Experimental Setup} 
We adopt SwinJSCC~\cite{yang2024swinjscc} \footnote{We select base SwinJSCC model without rate-adaptation and SNR-adaptation module.} as the representative transformer-based SemCom system for our analysis. 
The model is trained on the COCO dataset \cite{lin2014microsoft}, and tested on the Kodak dataset \footnote{The training and testing datasets are consistent with those used in Section~\ref{cnn-analysis} and Section~\ref{exp}. 
Detailed descriptions will be provided in Section~\ref{exp}. Besides, the test dataset can be found at http://r0k.us/graphics/kodak/}. 
The detailed parameters of the semantic encoder are listed in Table.~\ref{swinjscc_para}. The channel bandwidth ratio (CBR) \footnote{The CBR refers to the ratio between the length of the transmitted semantic features and that of the original input data.} of is 0.125. All transformer-based architectures in this paper adopt the same CBR setting for a fair comparison.

\begin{table}[htbp]
\centering
\caption{SwinJSCC encoder networks, taking an image $ x \in \mathbb{R}^{3 \times 192 \times 256}$ as an example.}
\label{swinjscc_para}
\begin{tabular}{l|c|l}
\toprule[1pt]
\textbf{Layer} & \textbf{Output Size}     & \textbf{Notes}      \\ \midrule[0.3pt]
Patch Embedding         & $12288 \times 128$ &  \\ 
Swin Transformer Block * 2              & $12288 \times 128$    &    Num\_heads = 4   \\ 
Patch Merging                   & $3072 \times 192$           &     DownSample       \\ 
Swin Transformer Block * 2                    &   $3072 \times 192$ &  Num\_heads = 6                    \\
Patch Merging                   & $768 \times 256$            &    DownSample       \\ 
Swin Transformer Block * 4                    &    $768 \times 256$     &   Num\_heads = 8            \\
Patch Merging                   & $192 \times 256$         &      DownSample        \\ 
Swin Transformer Block * 2                    &   $192 \times 256$      &    Num\_heads = 8         \\   
Norm \& FC                    &    $192 \times 96$         &         \\   \midrule[0.5pt]
Total Trainable Parameters & \multicolumn{2}{c}{7.70 M} \\ \bottomrule[1pt]
\end{tabular}
\end{table}
\paragraph{Inter-channel Partitioning Test}
To evaluate the robustness of semantic tokens against partial loss within a single token, we design an inter-channel partitioning test. Specifically, instead of dropping entire tokens, we simulate packet loss erasures by randomly masking portions of the semantic feature output channels.

\begin{figure*}[h]
    \centering
    \includegraphics[width=0.95\textwidth]{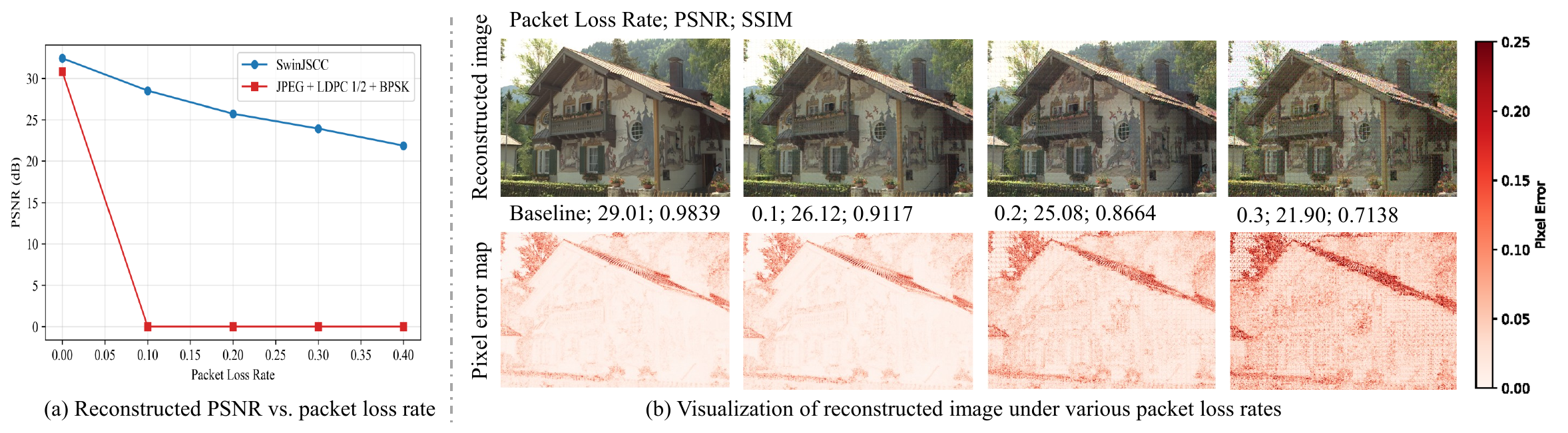}
    \caption{Performance of inter-channel partitioning strategy: PSNR curve and visualization. }
    \label{fig:intra_tr_test}
\end{figure*}

Fig.~\ref{fig:intra_tr_test}~(a) illustrates the degradation trend. Although the PSNR decreases as the erased ratio increases, the system maintains reasonable reconstruction quality even when 40\% of channels within each token are dropped.

Fig.~\ref{fig:intra_tr_test}~(b) further provides a pixel-level visualization of reconstruction errors under inter-channel erasure.
The error maps reveal that most regions remain well preserved, with noticeable deviations only appearing around object boundaries and fine texture areas.
This indicates that the semantic information within each token is evenly distributed across embedding dimensions, allowing the model to compensate for partial feature loss through the remaining components.

These results indicate that when semantic features are split along the channel dimension, the model exhibits inherent resilience against packet loss. This may stem from information being dispersed across multiple channels rather than concentrated in specific dimensions. Whether this property is broadly present across models or specific to the channel dimension requires further validation through comparative experiments involving token dimension splitting.

\paragraph{Inter-token Partitioning Test}
While the previous experiment focuses on feature partitioning within a single channel, we next examine the effect of partitioning at the token level. In this setting, semantic features are divided into discrete tokens, and entire tokens are randomly dropped during transmission to simulate packet loss at the granularity of semantic units. This experiment aims to evaluate how token-level partitioning influences the system’s robustness.

\begin{table}[htbp]
\centering
\caption{PSNR performances on inter-token test. $\Delta$PSNR = PSNR with loss $-$ PSNR without loss.}
\label{inter-token test}
\begin{tabular}{c|c|c}
\toprule[1pt]
\textbf{Packet Loss Rate} & \textbf{Avg PSNR(dB)}     & \textbf{$\Delta$PSNR(dB)}      \\ \midrule[0.3pt]
0.1        & 24.65 &  -7.79\\ 
0.2             & 21.56   &  -10.88     \\ 
0.3                   & 19.83          &    -12.61       \\    \bottomrule[1pt]
\end{tabular}
\end{table}
As shown in Table.~\ref{inter-token test}, the reconstruction images experience a noticeable degradation in PSNR compared with the case without packet loss.

\begin{figure*}[h]
  \centering
  \begin{subfigure}[b]{0.25\textwidth}
    \includegraphics[width=\linewidth]{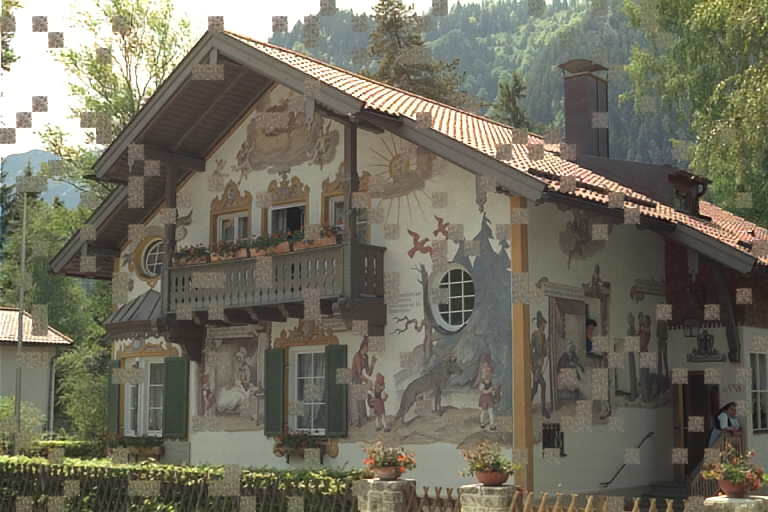}
    \caption{$p=0.1$}
  \end{subfigure}
  \hspace{1em}
  \begin{subfigure}[b]{0.25\textwidth}
    \includegraphics[width=\linewidth]{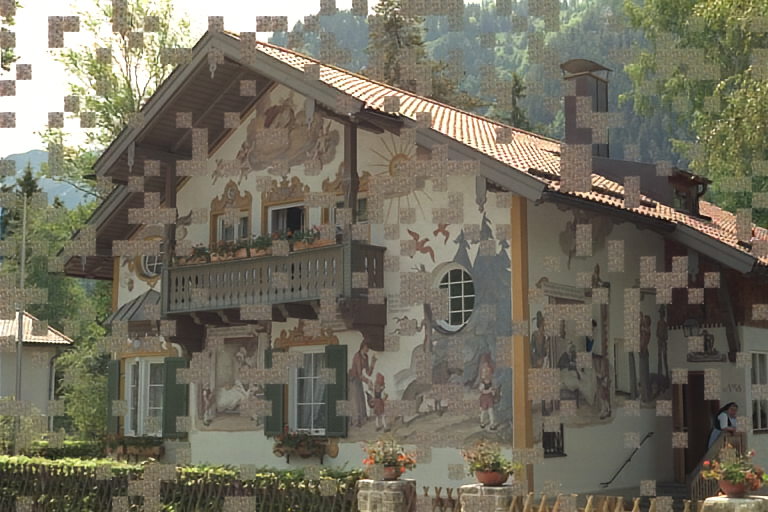}
    \caption{$p=0.2$}
  \end{subfigure}
  \hspace{1em}
  \begin{subfigure}[b]{0.25\textwidth}
    \includegraphics[width=\linewidth]{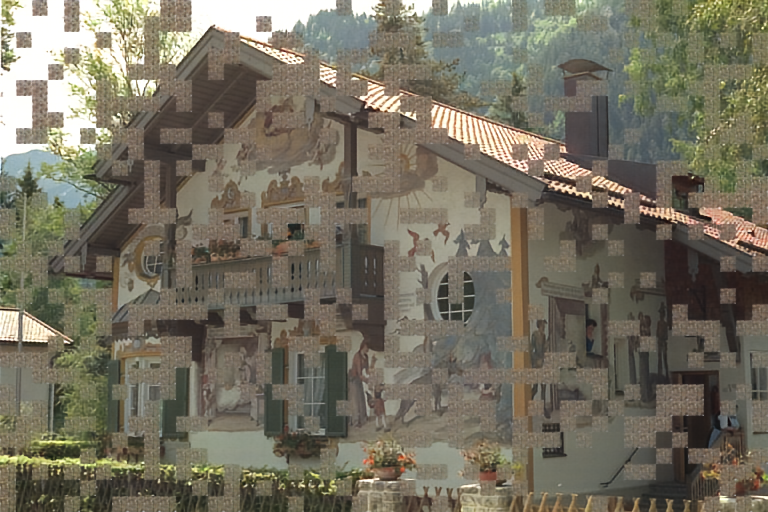}
    \caption{$p=0.3$}
  \end{subfigure}
  \caption{Visualization of the impact of different inter-token loss rates on reconstructed images.}
  \label{fig:inter_token_vis}
\end{figure*}
To further investigate the impact of entire token loss, we conduct a qualitative analysis by visualizing reconstructed images under different loss ratios. As shown in Fig.~\ref{fig:inter_token_vis}, masking an entire token leads to localized black patches at the corresponding spatial positions in the reconstructed image. Moreover, these missing regions often induce artifacts in the surrounding areas, since the decoder attempts to interpolate missing semantics based on incomplete contextual information.

This phenomenon indicates that the degradation caused by the loss of an entire token cannot be compensated through inter-token dependencies, despite the attention-based interactions among tokens.
In other words, the inherent token-to-token relationships are insufficient to counteract the impact of packet loss.

\subsection{Feature Partitioning in CNNs-based Semantic Communication Systems}
\label{cnn-analysis}

\paragraph{Model Design and Implementation} 

\begin{figure*}[ht]
    \centering
    \includegraphics[width=0.98\textwidth]{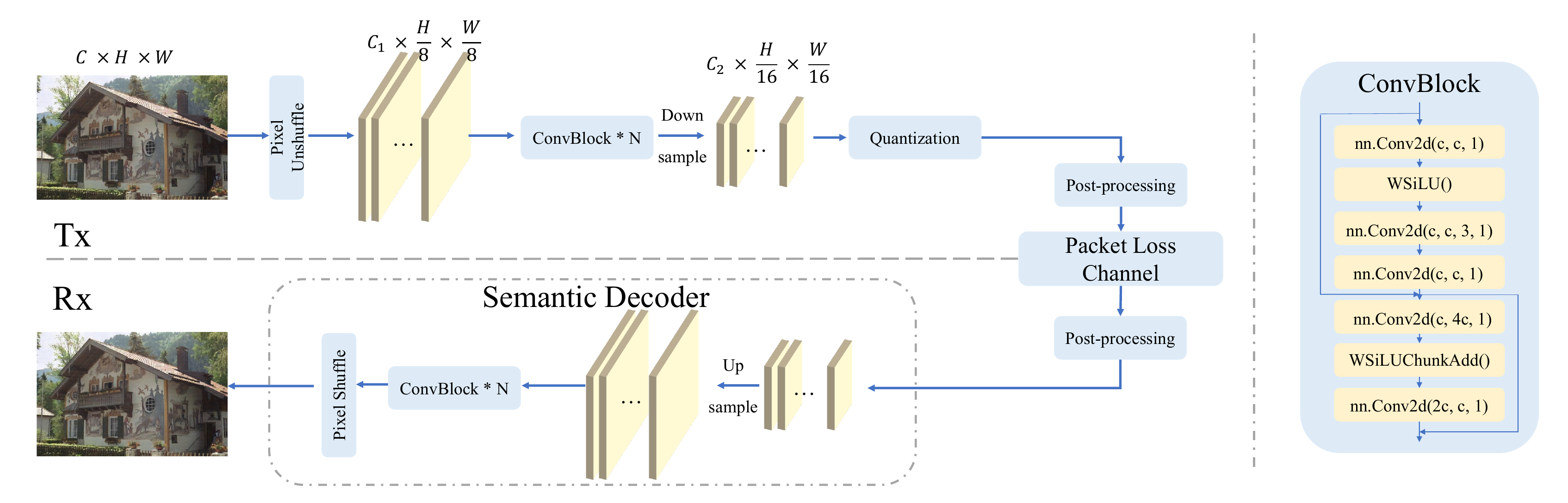}
    \caption{CNNs-based image semantic communication system framework.}
    \label{cnn_model}
\end{figure*}
The CNNs-based baseline model follows a structure similar to that in~\cite{ipc2019TCCN}, where we adopt the image semantic encoder/decoder proposed in~\cite{jia2025towards} with several modifications. In particular, to align the compression standard with that of SwinJSCC, we remove components such as entropy coding that are not directly related to semantic representation learning. These modifications allow us to focus purely on the semantic-level robustness, without the influence of additional coding mechanisms. The overall architecture is illustrated in Fig.~\ref{cnn_model}. The CBR is 0.333.

\paragraph{Inter-channel Partitioning Test}
To further examine the robustness of CNNs-based representations, we conduct an intra-channel partitioning test.
This experiment aims to evaluate how resilient the semantic features are to partial loss within individual channels.
Specially, we simulate inter-channel corruption by randomly erasing different proportions of feature elements along the channel dimension.

\begin{figure}[htbp]
    \centering
    \includegraphics[width=0.6\linewidth]{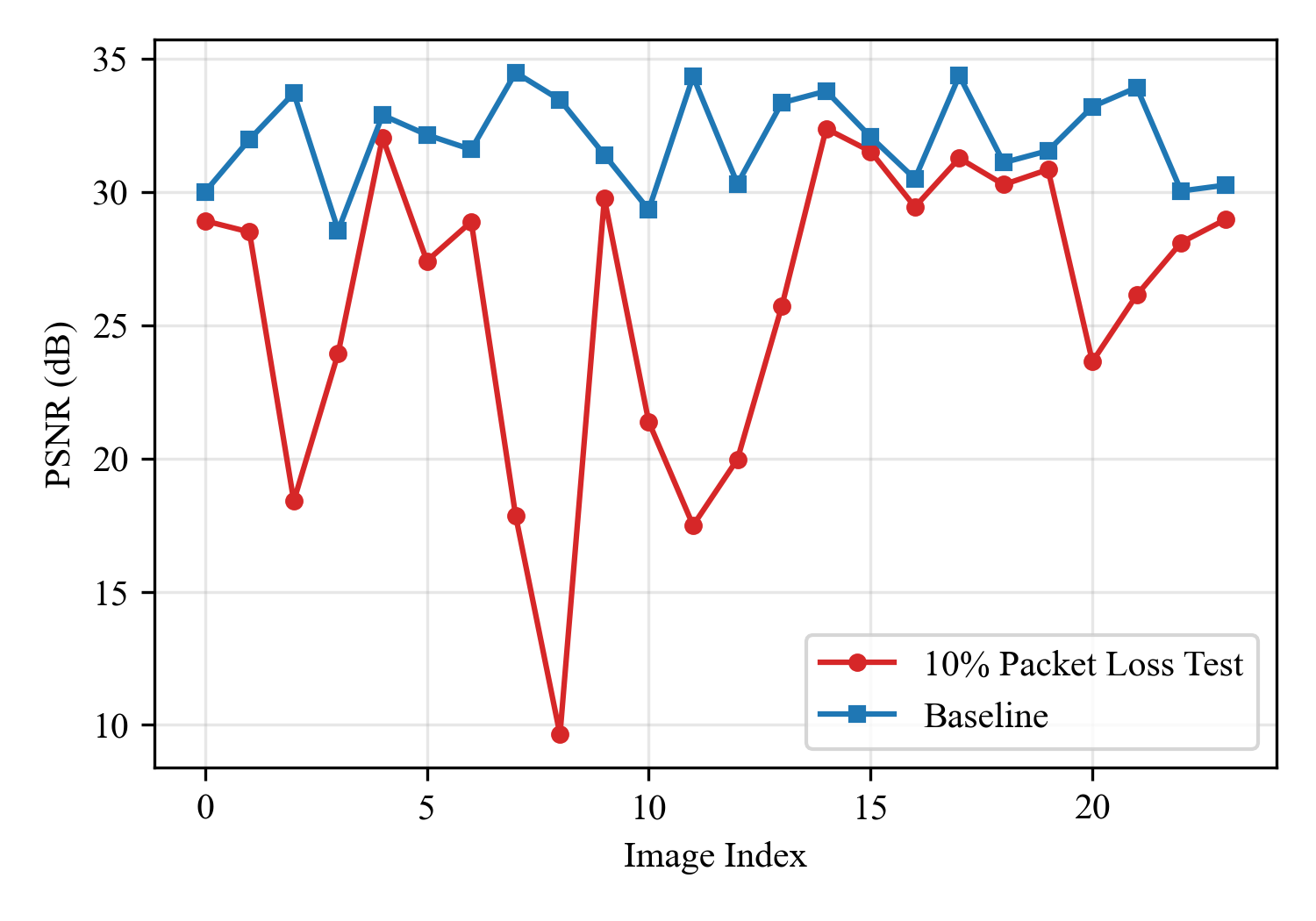}
    \caption{Per-image PSNR on Kodak under packet loss—baseline versus cross-channel partitioning.}
    \label{fig:channel_cnn_test}
\end{figure}

Experimental results show that, on average, the reconstructed PSNR of the proposed model under 10\% packet loss conditions decreases by approximately 10dB compared with the lossless case. However, when analyzing the reconstruction quality of individual images, the performance degradation exhibits a large variance, indicating that while the overall trend is a noticeable decline, the sensitivity to packet loss differs significantly across samples. Detailed $\Delta$PSNR can be found in Fig.~\ref{fig:channel_cnn_test}.

The results in Fig.~\ref{fig:channel_cnn_test} reveal two important observations.
First, the reconstruction quality remains relatively high for many images even when a portion of channels is removed, indicating that channel-wise partitioning can tolerate partial loss due to the presence of distributed semantic information.
Second, the reconstruction performance varies notably across different images, implying that the semantic information is unevenly distributed among channels.
When channels carrying critical semantics are excluded, the reconstruction quality degrades sharply.

\begin{figure}[htbp]
    \centering
    \includegraphics[width=0.6\linewidth]{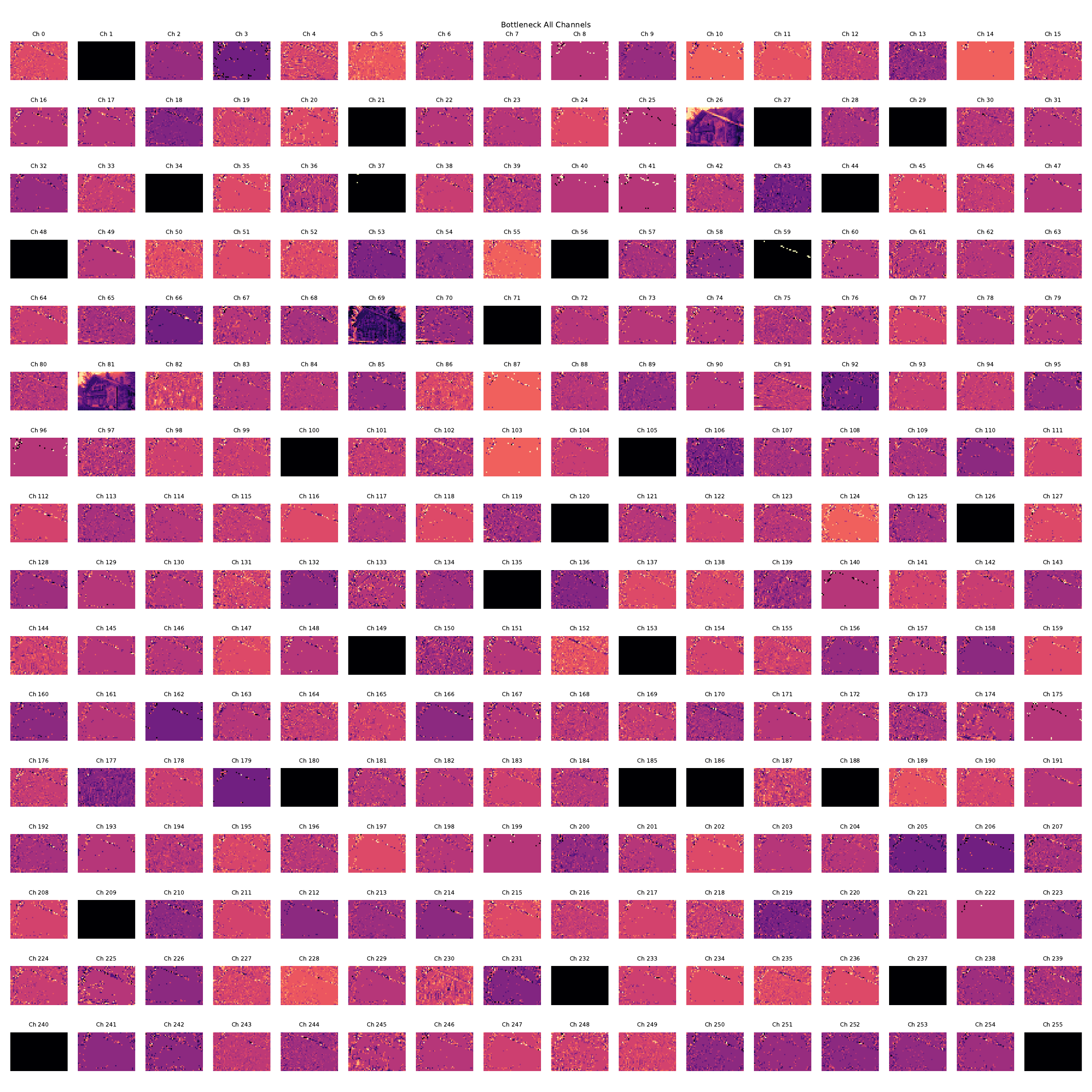}
    \caption{Channel-wise visualization of the CNNs-based semantic encoder outputs.}
    \label{fig:channel_vis}
\end{figure}
To further validate this observation, we visualize the feature maps at the bottleneck layer of the encoder, as Fig.~\ref{fig:channel_vis}. The visualization reveals that certain channels dominate the representation of key semantic content, like channel 26 and 69, confirming the non-uniformity of channel importance.  

The results of the channel-wise partitioning test reveal that the semantic information is distributed highly unevenly among them.
When the latent feature is partitioned along the channel dimension, some channels carry dominant semantic information that plays a decisive role in reconstruction quality, whereas others contribute only marginally.
Such imbalance in semantic importance makes the system particularly vulnerable when packets corresponding to critical channels are lost, even though redundant channels remain available.
These observations suggest that channel-wise partitioning alone cannot ensure robustness against packet loss unless the semantic contribution is balanced across channels.
If the information distribution across CNNs channels could achieve a similar level of balance as that observed within transformer tokens, the system would likely exhibit smoother performance degradation and stronger resilience under lossy transmission.

\paragraph{Inter-spatial Partitioning Test}
To examine whether spatial redundancy within feature maps can compensate for transmission errors, we conduct an inter-spatial partitioning test.
In this experiment, feature maps are partially corrupted by randomly masking specific spatial locations across all channels, simulating packet loss that removes information from localized positions while preserving the overall channel structure.

The reconstructed results show a drastic degradation in image quality, with PSNR dropping sharply even at moderate erasure ratios. The reconstructed results are analogous to the case in transformer architecture where entire tokens are lost: once a spatial region is masked, the information contained in that region cannot be effectively reconstructed, indicating that the model lacks sufficient spatial redundancy to infer the missing content.

Partitioning along the spatial dimension exhibits inferior performance compared with channel-wise segmentation and is therefore less effective as a partitioning strategy in CNNs-based SemCom systems.

\subsection{Discussion}
From the above experiments, it can be observed that the choice of feature partitioning dimension has a significant impact on the model’s robustness against packet loss. Notably, both Transformer- and CNNs-based SemCom systems perform better when features are partitioned along the channel dimension~\footnote{This observation motivates the choice of channel-wise partitioning as the default strategy in the following experiments.}. This may be attributed to the fact that each channel often represents a holistic or semantically coherent aspect of the image, making channel-based segmentation more aligned with the intrinsic structure of the encoded representation.

However, the two architectures exhibit distinct semantic distribution characteristics. Transformers tend to encode information more uniformly across tokens due to their global attention mechanism, while CNNs show an uneven distribution of channel importance, where certain channels dominate semantic reconstruction.

Furthermore, these findings can also be interpreted from the perspective of semantic redundancy—the fact that some features can be removed or altered with minimal impact on reconstruction performance. This property suggests potential for applying compression-oriented techniques, such as feature fusion or pruning, to further improve bandwidth efficiency without significantly compromising task performance.
\section{Semantic Equalization Mechanism}
This section introduces the motivation of semantic equalization mechanism, followed by two implementation methods: dynamic scaling and channel broadcasting. The subsequent parts describe the overall training process and provide discussions on the potential applications in task-oriented communication.
\subsection{Motivation}
Through our analysis of CNNs-based semantic features, we observe that the importance of channels is highly uneven: some channels carry critical information for reconstruction, while others contribute marginally. When important channels are lost during transmission, reconstruction quality, measured by PSNR, drops significantly, whereas loss of less important channels has minor impact. This suggests that balancing the contribution of each channel could mitigate the adverse effects of packet loss, effectively making the system more robust. Inspired by this observation, we propose the concept of \emph{Semantic Equalization}, which aims to regulate the influence of each channel, ensuring that no single feature dominates the reconstruction process. This notion is aligned with the classical idea of equalization in communication systems—where signal components are adjusted to achieve comparable strength—but here we extend it to the semantic domain by encouraging a more uniform contribution across latent semantic channels.

This motivation naturally leads to two practical implementations of semantic equalization.
\subsection{Dynamic Scale Module Inspired by Dynamic Neural Networks}
\subsubsection{Preliminaries on Dynamic Neural Networks}
Dynamic neural networks (DyNNs) can adapt their architectures or computation paths according to the input, thereby improving efficiency through selective execution. For example, several prior works \cite{wang2018skipnet, Wu2017BlockDropDI} introduce learnable gating mechanisms that dynamically skip layers or prune channels in response to the inputs. Extending this idea beyond conventional vision tasks, \cite{shao2023task} employs DyNNs to perform variable-length feature encoding, demonstrating their potential for communication-oriented scenarios.

In this work, we draw inspiration from the concept of selective activation in DyNNs. Instead of optimizing computational efficiency, our objective is to dynamically adjust the activation strength of semantic features based on their relative importance, thereby enhancing the robustness of semantic communication systems.
\subsubsection{Mathematical Formulation}
To efficiently balance the contribution of different channels, we introduce a \emph{semantic importance vector} 
$\boldsymbol{\gamma}$ to represent the relative importance of each channel dimension in the encoded feature space. 
Considering the encoder output $Z \in \mathbb{R}^{c \times h \times w}$, where $c$ denotes the number of channels, 
we model the semantic transformation through a weighted mapping function:

\begin{equation*}
S(\mathbf{z}) = \mathbf{W}\mathbf{z} + \mathbf{b} = \tilde{\mathbf{W}}\tilde{\mathbf{z}},
\end{equation*}
where $\tilde{\mathbf{W}} = [\mathbf{W}, \mathbf{b}]$ is the augmented weight matrix and 
$\tilde{\mathbf{z}} = [\mathbf{z}^T, 1]^T$ is the augmented input vector.
Denoting the $i$-th channel’s weight vector in $\tilde{\mathbf{W}}$ as $\tilde{\mathbf{W}}_i$, 
and the corresponding element in $\boldsymbol{\gamma}$ as $\gamma_i$, we rewrite the augmented weight matrix as

\begin{equation*}
\hat{\mathbf{W}}_i = \gamma_i \frac{\tilde{\mathbf{W}}_i}{\|\tilde{\mathbf{W}}_i\|_2},
\end{equation*}
where $\gamma_i$ serves as a dynamic scale factor that regulates the contribution of the $i$-th channel.
The mapping from the input feature $\mathbf{x}$ to the scaled semantic representation $\mathbf{z}$ can thus be formulated as

\begin{equation*}
\mathcal{z}_i = \tanh \left( \gamma_i \frac{\tilde{\mathbf{W}}_i}{\|\tilde{\mathbf{W}}_i\|_2} f(\mathbf{z}) \right),
\end{equation*}
where $\mathcal{z}_i$ is the output corresponding to the $i$-th semantic unit, and $\tanh(\cdot)$ is the activation function.
Here, the scaling factor $\gamma_i$ is designed to balance semantic importance across dimensions, 
preventing over-dominance of specific units and improving robustness under packet loss. 
This dynamic scaling module allows the network to adaptively distribute semantic information 
such that the reconstruction degrades gracefully when certain units are lost during transmission.

\subsubsection{Complexity Analysis}
The dynamic scale method generates a scaling vector that adaptively adjusts the importance of semantic channels according to the current channel condition.

Instead of using fixed or handcrafted scaling coefficients, the scaling factors are learned through a lightweight neural network, denoted as $\Gamma(\cdot)$, which outputs a set of normalized scaling values.
This network is composed of four fully connected layers with progressively increasing dimensionality, as specified below.

For the four-layer mapping $\Gamma(\cdot)$, the total number of parameters is
\[
\#\text{Param} = \sum_{i=1}^{4} (d_i \times d_{i-1} + d_i),
\]
where $(d_0, d_1, d_2, d_3, d_4) = (1, 16, 16, 16, c)$. 
Hence,
\[
\begin{aligned}
\#\text{Param} &= (16 \times 1 + 16) + (16 \times 16 + 16) \times 2 + (c \times 16 + c) \\
               &= 576 + 17C.
\end{aligned}
\]
For $c=256$, this method introduces approximately $4.9 \times 10^3$ parameters, which is negligible compared to the encoder backbone.

The computational complexity is dominated by the matrix-vector multiplications in $\Gamma(\cdot)$ and the linear transformation $\mathbf{U}\boldsymbol{\mu}$.
The total FLOPs can be expressed as:
\[
\text{FLOPs} = \sum_{i=1}^{4} d_i d_{i-1}.
\]
For $c=256$, this yields approximately $4.6 \times 10^3$ FLOPs, which is below $0.5\%$ of a typical transformer or CNNs encoder stage.
Thus, this method maintains extremely low computational overhead and is suitable for real-time deployment.
\subsection{Broadcast Module for Semantic Diffusion}

While the dynamic scale mechanism focuses on rebalancing the contribution of each semantic unit individually, the broadcast mechanism aims to redistribute semantic information across units through semantic interactions. Together, these two approaches form a unified semantic equalization framework.

Specifically, inspired by the message-passing paradigm in graph neural networks (GNNs)~\cite{shen2020graph, atwood2016diffusion}, we treat the feature channels as nodes in a semantic topology, where information can be locally propagated along predefined edges. Instead of performing global mixing, we introduce a neighboring semantic broadcast (NSB) idea that diffuses semantic information only to topologically nearest $K$ neighbors, thereby enabling neighboring channels to share partially overlapping semantics and compensating for packet loss.

Formally, given the encoded feature tensor $\mathbf{Z} \in \mathbb{R}^{c \times h \times w}$, we define a sparse and normalized broadcast matrix $\tilde{\mathbf{B}} \in \mathbb{R}^{c \times c}$ representing the adjacency between neighboring channels. Each element $\tilde{B}_{ij}$ represents the diffusion weight from channel $j$ to channel $i$, defined as:
$$
\tilde{B}_{i j}= \begin{cases}\frac{1}{K}, & \text { if } j \in \mathcal{N}_K(i), \\ 0, & \text { otherwise }\end{cases}
$$
where $\mathcal{N}_K(i)$ denotes the set of the $K$ nearest neighbors of channel $i$ in the semantic topology. The diffusion process is then formulated as:
$$
\tilde{Z} = \tilde{B}Z,
$$
where the operation is independently applied to each spatial location $(h, w)$.
This allows each channel to aggregate complementary information from its $K$ nearest neighbors, analogous to localized message passing in GNNs, forming a low-cost semantic equilibrium across the channel topology.

The NSB maintains local redundancy and structural fidelity, and its localized broadcasting enables richer and more diverse semantic representations than global broadcasting. Moreover, since the broadcast matrix is predefined and non-learnable, the mechanism introduces no additional parameters or training overhead, making it easily integrable into existing semantic encoders.

\subsection{Training Procedure}
\begin{algorithm}[t]
\caption{Training Process}
\label{training}
\begin{algorithmic}[1]  

\Require Image dataset, semantic encoder $E_\theta(\cdot)$, semantic decoder $D_\psi(\cdot)$, semantic equalization module $\mathrm{SEM}(\cdot)$
\Ensure Trained $E_\theta$, $D_\psi$, and optionally SEM
\While{stop criterion not met}
    \State Encode input images: $\mathbf{F} = E_\theta(\mathbf{I})$
    \State Apply semantic equalization: $\mathbf{F}_{\mathrm{eq}} = \mathrm{SEM}(\mathbf{F})$
    \State Quantize features: $\mathbf{Z} = Q(\mathbf{F}_{\mathrm{eq}})$
    \State Transmit $\mathbf{Z}$ through the (simulated) channel (no packet loss during training); 
receive $\hat{\mathbf{Z}}$
    \State Decode features: $\hat{\mathbf{F}}_{\mathrm{eq}} = D_\psi(\hat{\mathbf{Z}})$
    \State Reconstruct images: $\hat{\mathbf{I}} = \hat{\mathbf{F}}_{\mathrm{eq}}$ 
    \State Compute MSE loss: $\mathcal{L} = \|\mathbf{I} - \hat{\mathbf{I}}\|^2$
    \State Update parameters $\theta, \psi$ (and optionally SEM) to minimize $\mathcal{L}$
\EndWhile
\end{algorithmic}
\end{algorithm}
The proposed SEM can be seamlessly integrated after the semantic encoder and jointly trained with the entire system in an end-to-end manner, without the need for additional pre-training or fine-tuning. More implementation details are provided in Algorithm.~\ref{training}.
\subsection{Dicussion}
Beyond reconstruction-oriented semantic communication, the SEM could potentially extend to task-oriented communication scenarios.
In such systems, the encoder tends to prioritize task-relevant semantics—those most closely related to downstream objectives (e.g., classification logits or detection cues)—while less relevant details may be underrepresented.
This may lead to another form of semantic imbalance~\cite{shwartz2017opening, xie2023robust}, where task-relevant units dominate the feature space.
Based on this observation, it is reasonable to hypothesize that applying SEM in task-oriented settings could help moderate the importance distribution across task-specific features, thereby reducing the risk of over-concentration on a few highly sensitive dimensions~\cite{shao2021learning, shao2023task}.
Similar to prior works that explicitly adjust feature distributions through loss-function regularization~\cite{xie2023robust, zhang2025spectral}, SEM introduces a comparable balancing effect at the network level, enabling adaptive reallocation of representational capacity among semantic components.
Accordingly, SEM might serve as a general strategy for balancing semantic representations in both reconstruction-based and task-driven communication systems, potentially enhancing robustness under imperfect transmission conditions.

\section{EXPERIMENTS RESULTS}
In this section, numerical results are provided to validate the effectiveness of the proposed SEM. Besides, we offer a perspective for understanding this intrinsic robustness.
\label{exp}

\subsection{Experimental Setups}
\subsubsection{Datasets}
For training, we adopt the COCO 2014 dataset \cite{lin2014microsoft}, which contains over 80k training images from diverse real-world scenes with rich object categories, textures, and complex backgrounds. This large-scale dataset provides sufficient variability to ensure the generalization capability of semantic communication models. For evaluation, we employ the Kodak dataset, consisting of 24 uncompressed high-quality images of size 768×512. Additionally, we use the DIV2K dataset~\cite{Agustsson_2017_CVPR_Workshops}, which contains 100 high-resolution RGB images (800–1200 pixels in the shorter side) provided in PNG format without any compression artifacts.

\subsubsection{Metric}
We quantify the image transmission performance with the following two metrics: pixel-wise metric PSNR, perceptual metric SSIM. In this work, PSNR is computed by $$PSNR = 10\log_{10}(\frac{MAX^2}{MSE}), $$where $MSE$ is mean squared error between the reconstructed and reference images and $MAX = 255$.
SSIM is computed using the implementation provided in the PyTorch Image Quality (piq) library. Overall, higher PSNR/SSIM indicates better performance.

\subsubsection{Model Deployment Details}
All models are implemented in PyTorch and trained on NVIDIA V100 GPUs. 
For fair comparison, we maintain identical backbone configurations and training hyperparameters across all baselines. 
\subsection{Experimental Results}

\begin{figure*}[h]
  \centering
  \begin{subfigure}[b]{0.32\textwidth}
    \includegraphics[width=\linewidth]{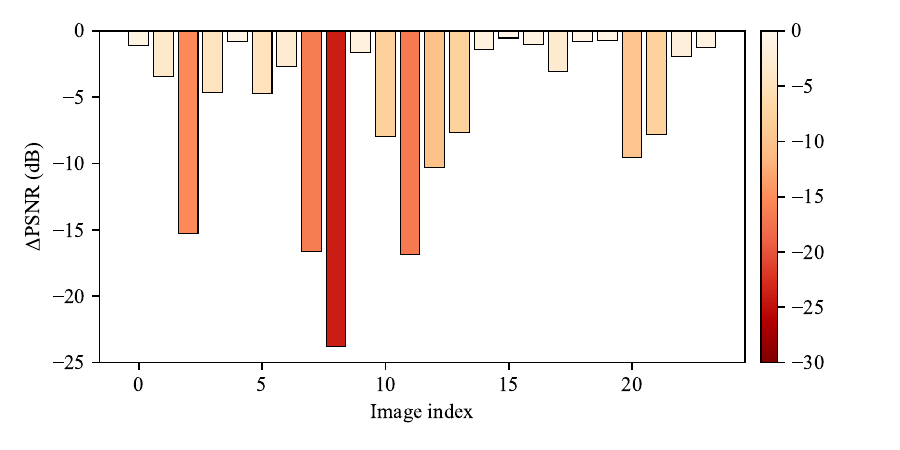}
    \caption{Baseline}
  \end{subfigure}
  \begin{subfigure}[b]{0.32\textwidth}
    \includegraphics[width=\linewidth]{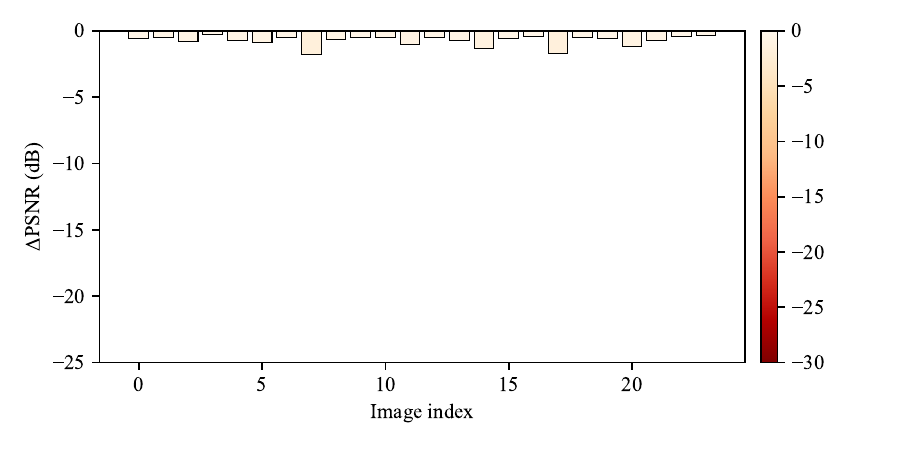}
    \caption{+ Scale}
  \end{subfigure}
  \begin{subfigure}[b]{0.32\textwidth}
    \includegraphics[width=\linewidth]{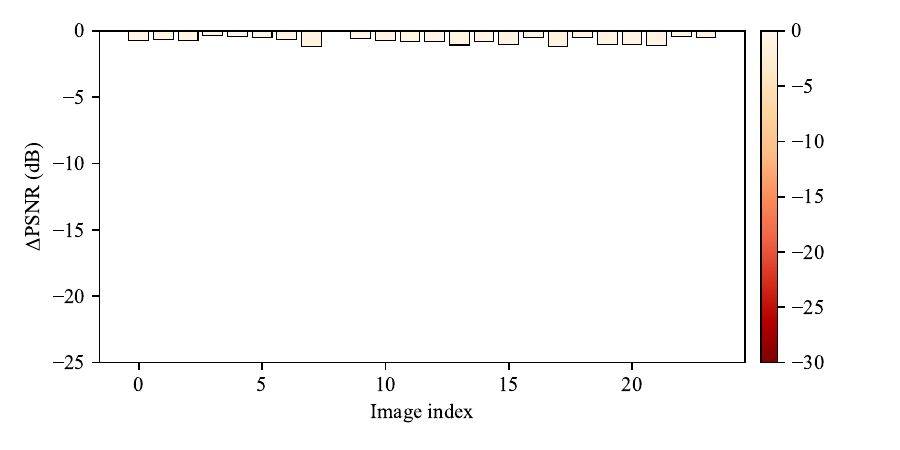}
    \caption{+ Broadcast}
  \end{subfigure}
  \caption{Per-image $\Delta$PSNR on the Kodak dataset under 10\% packet loss for CNNs-based frameworks. The “+Scale” variant represents the baseline enhanced with the proposed dynamic scaling mechanism, and “+Broadcast” denotes the addition of the NSB strategy.}
  \label{fig: bar_cnn_compare}
\end{figure*}

\subsubsection{Individual Reconstruction Quality} Fig.~\ref{fig: bar_cnn_compare} presents the per-image PSNR variations of different CNNs schemes on the Kodak dataset under 10\% packet loss. The baseline exhibits extremely large variance: some images suffer PSNR drops exceeding 20 dB, which severely degrades semantic-communication quality. After integrating the proposed SEM, the PSNR fluctuations become much more uniform, ensuring stable visual fidelity across all test images.

\begin{figure}[h]
    \centering
    \includegraphics[width=0.6\linewidth]{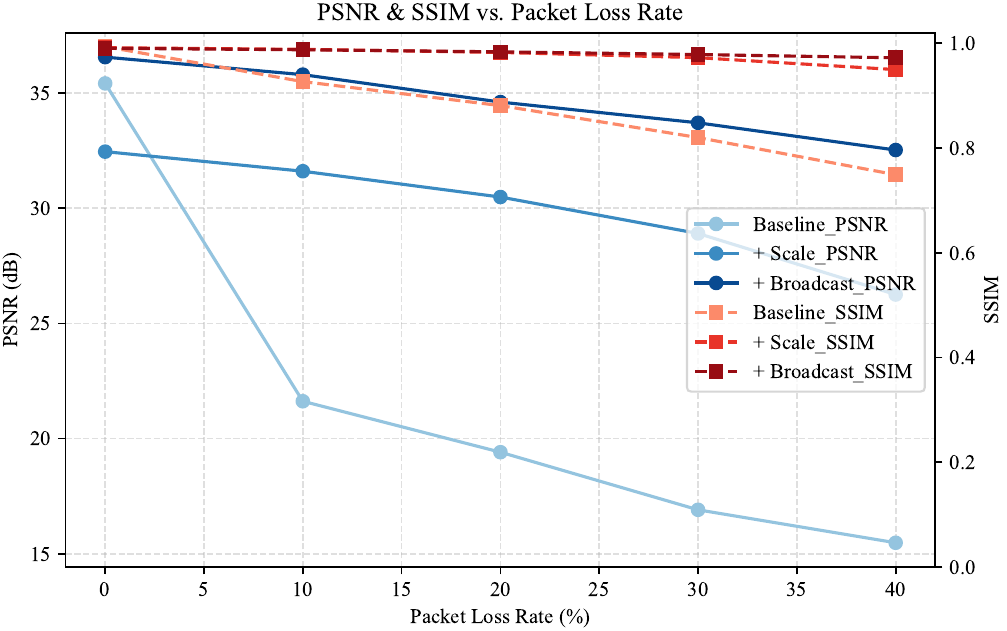}
    \caption{Average reconstruction quality vs. packet loss rate over the DIV2K Dataset for CNNs-based frameworks.}
    \label{total_performance}
\end{figure}
\subsubsection{Overall Performance} Fig.~\ref{total_performance} illustrates the average reconstruction quality of the proposed CNNs framework on the DIV2K dataset as a function of packet-loss rate.
Both +Scale and +Broadcast exhibit remarkable robustness against packet loss, as SSIM remains almost stable and PSNR only slightly decreases even under severe transmission degradation. Compared with the Baseline, the reconstructed PSNR increases by about 10 dB, highlighting the effectiveness of the proposed SEM.

\begin{figure*}[htbp]
    \centering
    \includegraphics[width=0.99\textwidth]{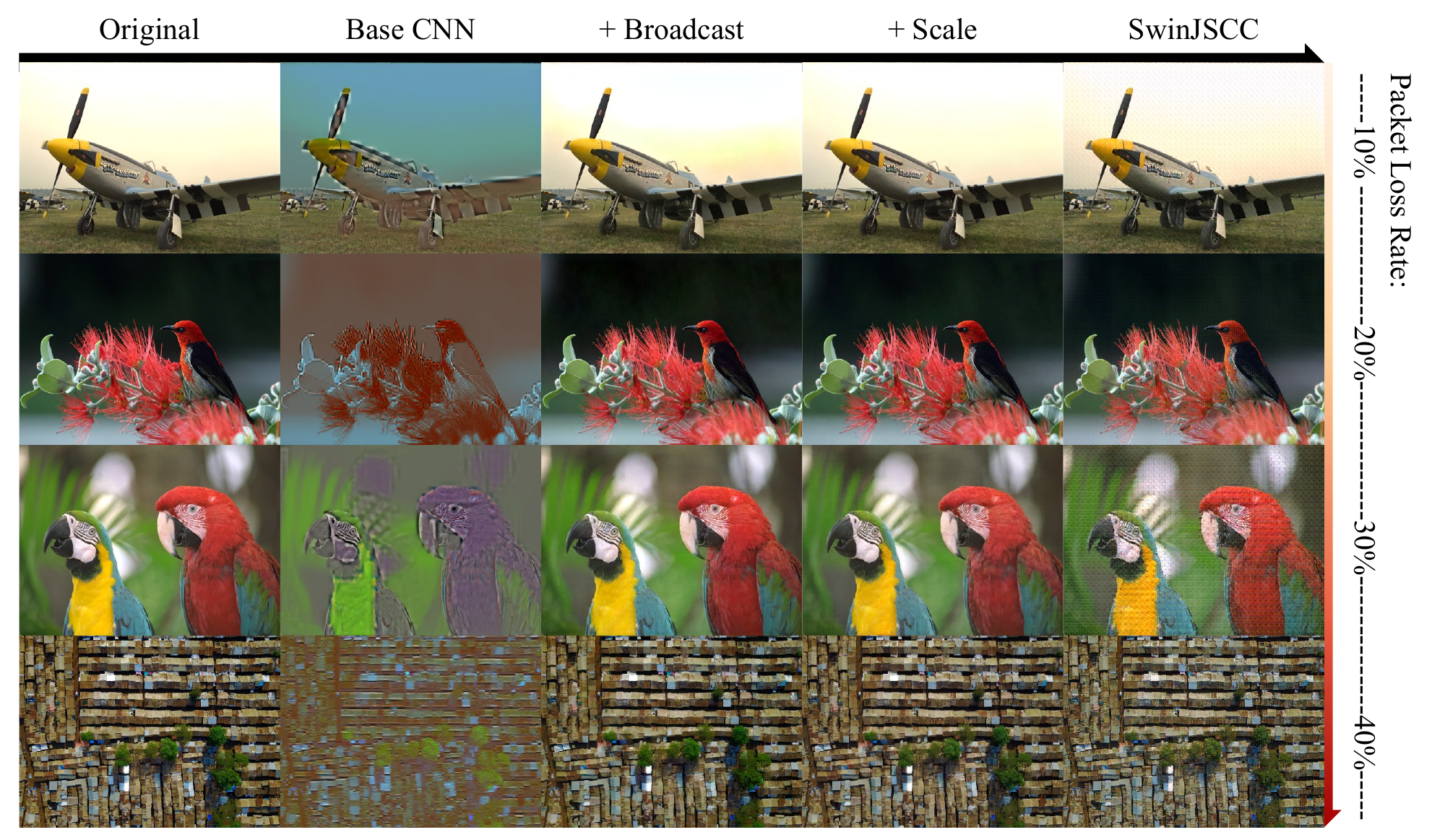}
    \caption{Visual comparison of reconstructed images under various packet-loss conditions. The results are provided for qualitative illustration only, without quantitative evaluation.}
    \label{fig: final_vis_compare}
\end{figure*}

\subsubsection{Visualization} Fig.~\ref{fig: final_vis_compare} presents visual comparisons of reconstructed images under packet loss conditions on the Kodak and DIV2K test sets. It can be clearly observed that the baseline CNNs-based scheme suffers from style shifts and noticeable blurring when important channels are lost. In contrast, the variants incorporating SEM effectively alleviate these issues. Meanwhile, SwinJSCC consistently maintains stable robustness against packet loss throughout all cases.

Overall, the above experiments verify the effectiveness of the proposed SEM in balancing channel importance, which effectively enables CNNs-based SemCom systems to achieve graceful performance degradation under packet loss conditions.

\subsection{One Perspective to Understand the Robustness Against Packet Loss: Don't Put All Your Eggs in One Basket}
\begin{figure*}[htbp]
  \centering
  \begin{subfigure}[b]{0.23\textwidth}
    \includegraphics[width=\linewidth]{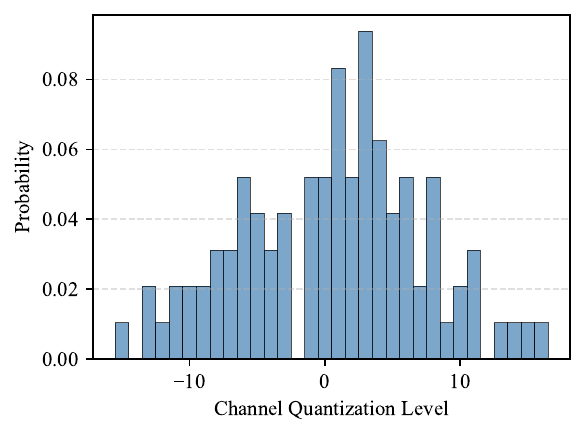}
    \caption{}
  \end{subfigure}
  \begin{subfigure}[b]{0.23\textwidth}
    \includegraphics[width=\linewidth]{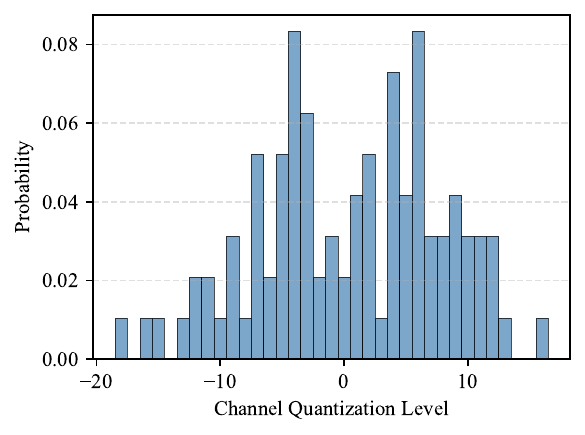}
    \caption{}
  \end{subfigure}
  \begin{subfigure}[b]{0.23\textwidth}
    \includegraphics[width=\linewidth]{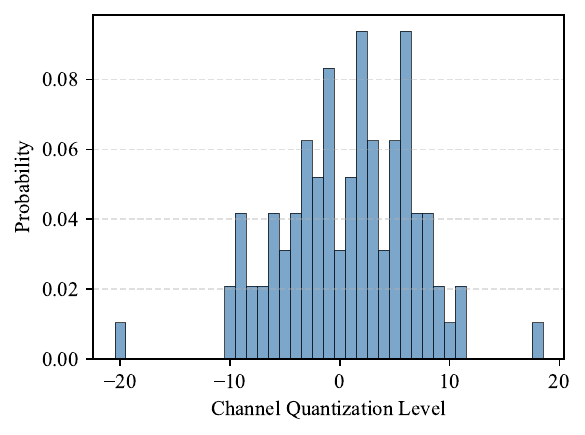}
    \caption{}
  \end{subfigure}
  \begin{subfigure}[b]{0.23\textwidth}
    \includegraphics[width=\linewidth]{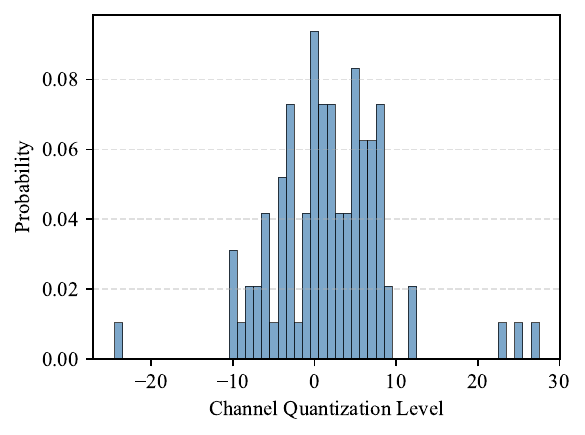}
    \caption{}
  \end{subfigure}
  \begin{subfigure}[b]{0.23\textwidth}
    \includegraphics[width=\linewidth]{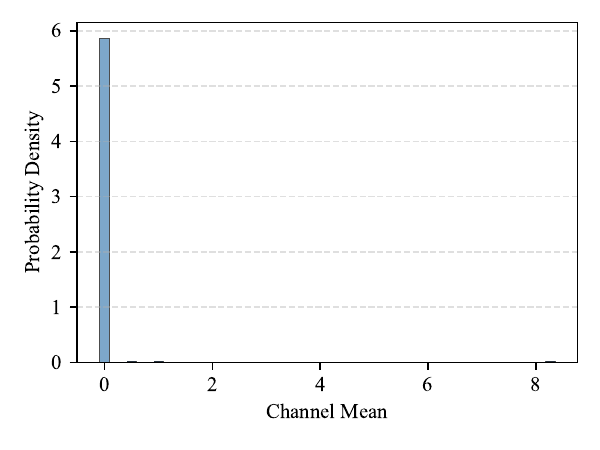}
    \caption{Baseline}
  \end{subfigure}
  \hspace{1.2em}
  \begin{subfigure}[b]{0.23\textwidth}
    \includegraphics[width=\linewidth]{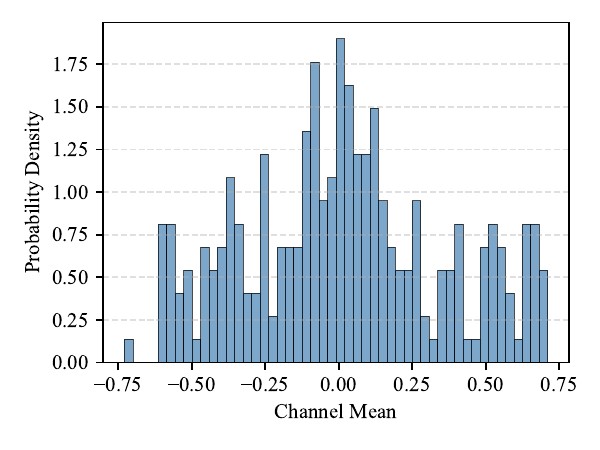}
    \caption{+ Scale}
  \end{subfigure}
  \hspace{1.2em}
  \begin{subfigure}[b]{0.23\textwidth}
    \includegraphics[width=\linewidth]{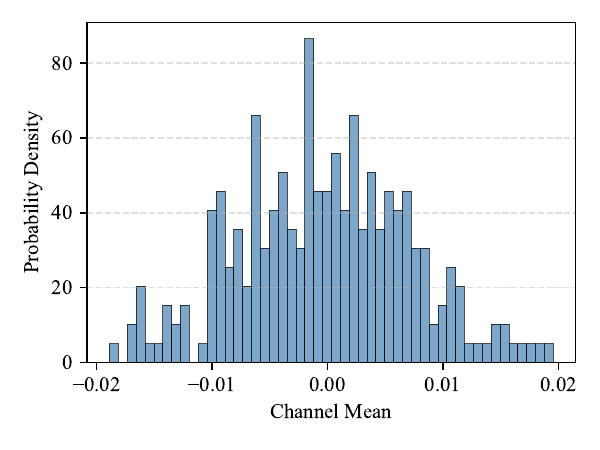}
    \caption{+ Broadcast}
  \end{subfigure}
  \caption{Probability distributions of different frameworks: (a)–(d) show the quantized value distributions of four randomly selected tokens in the Transformer-based architecture, while (e)–(g) depict the probability density distributions of channel means for the three CNNs-based schemes.}
  \label{fig:understanding_cnn}
\end{figure*}

To better understand the robustness of semantic equalization and the intrinsic properties of the transformer architecture under packet loss conditions, we plot the corresponding probability distributions, as shown in Fig.~\ref{fig:understanding_cnn}. For the transformer-based architecture, four tokens were randomly selected, and the occurrence probability of each quantized value within a token was plotted.
For the CNNs-based architecture, the mean value of each channel was calculated, and the corresponding probability density function was plotted.

From Fig.~\ref{fig:understanding_cnn}, we observe that in the baseline CNNs-based scheme, the channel means are mostly concentrated around zero.
In contrast, the schemes +Scale / Broadcast exhibit more uniformly distributed channel means.
Meanwhile, the transformer-based architecture, which is inherently robust to packet loss, also shows a relatively uniform distribution of token-level output values. The +Scale / Broadcast variants show a distribution pattern that is consistent with that observed in the transformer-based architecture.

From the perspective of entropy, a more uniform distribution inherently carries greater information content, as each feature or channel contributes more evenly to the overall representation.
\begin{equation*}
H(X) = - \sum p(x_i) \log p(x_i),
\end{equation*}

This increased entropy can be interpreted as an implicit form of robustness, enabling the semantic communication system to better withstand packet loss by avoiding over-reliance on a few highly sensitive dimensions. In practical scenarios, such a distribution characteristic enables a more effective use of the available quantization bits~\cite{gong2025digital}.

It should be noted, however, that such robustness is achieved at the expense of task-specific performance to some extent, reflecting an inherent trade-off between robustness and task optimization in semantic communication systems.

\section{Conclusion}

In this paper, we analyzed the performance of different feature partitioning strategies under packet loss, using both CNNs- and Transformer-based architectures for comparison. Experimental results show that Transformer models exhibit stronger robustness against packet loss when semantic features are partitioned along the channel dimension. In contrast, CNNs-based frameworks reveal imbalanced channel utilization, leading to degraded reconstruction quality once key features are lost. 

To address this issue, we introduced the SEM, including the Dynamic Scale and Neighboring Semantic Broadcast modules, which effectively rebalance semantic contributions and significantly improve reconstruction quality under various loss conditions.

Importantly, our study reveals a key principle: under proper feature partitioning, maintaining a balanced semantic representation—where different units contribute comparably to the task—is essential for achieving robustness against packet loss.

The findings and proposed mechanisms in this work may offer insights for practical deployment scenarios, and we expect that similar ideas could potentially generalize to other modalities such as video. Future work will focus on validating these approaches in real communication environments and exploring their integration into practical semantic transmission frameworks.


%




\ifCLASSOPTIONcaptionsoff
  \newpage
\fi



%
\small
\bibliographystyle{IEEEtran}
\bibliography{ref}

%








\end{document}